\documentclass[english,10.5pt,aps,prd,a4paper,preprintnumbers,floatfix,nofootinbib,showpacs,superscriptaddress, notitlepage]{revtex4-1} 
 \pdfoutput=1
\usepackage{color}
\usepackage{graphicx}
\usepackage{amsmath}
\usepackage{amssymb}
\usepackage{multirow}
\usepackage[colorlinks=true,citecolor=darkred,urlcolor=darkred, pdfborder={0 0 0}]{hyperref}
\usepackage[normalem]{ulem}
\definecolor{darkred}{rgb}{0,0.6,0}

\newcommand{\ba}{\begin{array}}
\newcommand{\ea}{\end{array}}
\def\be{\begin{equation}}
\def\ee{\end{equation}}
\def\bea{\begin{eqnarray}}
\def\eea{\end{eqnarray}}
\def\gsim{\ \rlap{\raise 2pt\hbox{$>$}}{\lower 2pt \hbox{$\sim$}}\ }
\def\lsim{\ \rlap{\raise 2pt\hbox{$<$}}{\lower 2pt \hbox{$\sim$}}\ }
\def\dslash{\kern-4pt \not{\hbox{\kern-2pt $\partial$}}}
\def\pslash{\not{\hbox{\kern-2pt p}}}

\begin{document}
\DeclareGraphicsExtensions{.eps,.ps}
%
\title{\boldmath Impact of RGE-induced $ \mu-\tau  $ Reflection Symmetry Breaking on the Effective Majorana Neutrino Mass in $ 0\nu\beta \beta $ Decay}
%
%

\author{Newton Nath}
\email[Email Address: ]{newton@ihep.ac.cn}
\affiliation{
Institute of High Energy Physics, Chinese Academy of Sciences, Beijing, 100049, China}
\affiliation{
School of Physical Sciences, University of Chinese Academy of Sciences, Beijing, 100049, China}



\begin{abstract}
{\noindent
We make an attempt to study the impact of renormalization-group equations (RGE) induced  $\mu-\tau$ reflection symmetry breaking on the effective Majorana neutrino mass $ |\langle m\rangle_{ee} | $ in  neutrinoless double beta ($ 0\nu\beta \beta $) decay. At present, the $ 0\nu\beta \beta $ decay serves as a unique process to address  the Majorana nature of massive neutrinos.
The rate of such decay  process depends on  $ |\langle m\rangle_{ee} | $. On the other hand, $\mu-\tau$ reflection symmetry predicts $ \theta_{23} = 45^\circ$ and $ \delta = \pm 90^\circ $  together with  trivial values of the Majorana CP-phases ($ \rho, \sigma $).  Moreover, based on the recent global best-fit values which prefer higher octant of $ \theta_{23} $ and third quadrant of $ \delta $,  it is hard to believe the exactness of such symmetry.
Also, any non-trivial values of $ \rho, \sigma $ may have some significant impact on $ |\langle m\rangle_{ee} | $. In this context, we study the spontaneous breaking of the symmetry via one-loop RGE-running from  a superhigh energy scale ($\Lambda_{\mu \tau} $)  down to the electroweak scale ($ \Lambda_{\rm EW} $). Given the broken symmetry, we perform some systematic analysis for $ |\langle m\rangle_{ee} | $ in substantial detail. Further, we also extend this analysis for other lepton-number violating effective Majorana masses. 
}
\end{abstract}

\maketitle

\section{Introduction}\label{sec:Intro}
Among a number of open questions in neutrino physics, the nature of neutrinos, whether they are Majorana or Dirac fermions, is yet unanswered \footnote{In 1937,  E. Majorana first hypothesized that a fermion can be its own antiparticle in Ref.~\cite{Majorana:1937vz}.}. At the current juncture, the only appreciable process which can uncover the Majorana nature of neutrinos is the neutrinoless double beta  ($ 0\nu\beta \beta $) decay~\cite{DellOro:2016tmg,Vergados:2016hso}. In this process, the following lepton-number violating decay takes place,
\begin{equation}
(Z, A) \longrightarrow (Z+2, A) + 2 e^{-} \;,
\end{equation}
 which offers a unique opportunity to test the violation of lepton-numbers by two-units \footnote{The testing of Majorana's theory was first suggested in Ref.\cite{Furry:1939qr}}.  In the standard three neutrino framework, the decay rate of such a process is controlled by the  effective Majorana neutrino mass $ |\langle m\rangle_{ee} | = \sum_i |m_i U^2{}_{ei}|$ where $ U_{e i} $'s (for $ i = 1, 2, 3 $) represent the elements of the first row of the Pontecorvo-Maki-Nakagawa-Sakata (PMNS) mixing matrix \cite{Patrignani:2016xqp}. Furthermore, the observation of the $ 0\nu\beta \beta $ decay process can also provide some information  about the absolute neutrino masses and constrain the Majorana CP-phases. At present, the most stringent upper bound on  $  |\langle m_{ee}\rangle| $ arises from KamLAND-Zen collaboration~\cite{KamLAND-Zen:2016pfg} and their latest data reports $  |\langle m_{ee}\rangle|  < (0.061 - 0.165)$ eV \footnote{Note that KamLAND-Zen collaboration gives lower bound for the  $ 0\nu\beta \beta $ decay half-life of $ T^{0\nu}_{1/2} $ at 90\% C.L. and considering  nuclear matrix element calculations, the
corresponding upper bound for the effective Majorana neutrino mass has been calculated.}, whereas the \textit{Planck} Collaboration \cite{Aghanim:2018eyx} gives bounds on the absolute neutrino mass scale as $ \sum m_{\nu} < 0.12$ eV (95\%, \textit{Planck} TT, TE, EE + lowE + lensing + BAO). 

On the other hand, the theory behind the dynamical origin of neutrino mass and leptonic flavor mixing yet needs to be answered. Among a large variety of theoretical models which address neutrino masses, the type-I seesaw mechanism~\cite{Minkowski:1977sc,Yanagida:1979as,GellMann:1980vs,Mohapatra:1979ia,Schechter:1980gr} has been considered the most simplest and elegant one. 
Furthermore, flavor symmetry has been very successful at explaining the observed leptonic mixing pattern  ~\cite{Altarelli:2010gt,Altarelli:2012ss,Smirnov:2011jv,Ishimori:2010au,King:2013eh}.
In this context, the  $\mu-\tau$ reflection symmetry, which was originally proposed in Ref.~\cite{Harrison:2002et}  (for the latest review, see Ref.~\cite{Xing:2015fdg} and the references therein) leads to $ |U_{\mu i }| = |U_{\tau i }|$, (for i = 1, 2, 3) where $ U $ is the PMNS flavor mixing matrix, and has received a great deal of attention. 
This symmetry predicts the maximal values of the atmospheric mixing angle $\theta_{23}^{} = 45^\circ$ as well as the Dirac CP-phase $\delta = \pm 90^\circ$ along with non-zero $ \theta_{13} $. 
 Moreover, it also predicts the trivial Majorana CP-phases $ \rho, \sigma = 0^\circ, 90^\circ $. Considering the latest global best-fit results 
which favors non-maximal $\theta_{23}^{}, \delta $~\cite{Capozzi:2016rtj,Esteban:2016qun,deSalas:2017kay}, it is hard to strict on the exactness of the symmetry.
Besides this, although the latest T2K~\cite{Abe:2017uxa} results are in good agreement with the concerned symmetry, the current NO$\nu$A~\cite{NOVA2018} results seem to favor non-maximal $\theta_{23}^{}, \delta$.
  On the other hand, as the flavor symmetries are generally imposed at a superhigh energy scale  to address tiny neutrino masses and their flavor mixing at low energies, the renormalization group equations (RGE) running
effect may lead to possible corrections and naturally break the exact symmetry. Indeed, in recent times breaking of  the  $\mu-\tau$ reflection symmetry via quantum corrections has received a great deal of attention~\cite{Zhao:2017yvw,Rodejohann:2017lre,Liu:2017frs,Xing:2017mkx,Xing:2017cwb,Nath:2018hjx,Huan:2018lzd,Zhu:2018dvj}. Recently, the impact of such symmetry for the upcoming long baseline neutrino oscillation experiment has been addressed in Ref.\cite{Nath:2018xkz}.
%

In this work, to the best of our knowledge for the first time, we make an attempt to study the impact of RGE-induced $\mu-\tau$ reflection symmetry breaking  on $ 0\nu\beta \beta $ decay. 
 We implement the concerned symmetry at the superhigh energy scale $ \Lambda_{\mu \tau} $ and 
analyze the RGE-triggered quantum corrections all the way from $ \Lambda_{\mu \tau} $ down to the electroweak (EW) scale $ \Lambda_{\rm EW} $.  Keeping the current global best-fit result in mind which prefers $ \theta_{23} > 45^\circ $ and $ \delta < 270^\circ $, we study the correlation among these two parameters due to RGE corrections at low energies.
 Further, the Majorana nature of massive neutrinos, which also suggests the existence of Majorana CP-phases, has some significant impact on  $ 0\nu\beta \beta $ decay.
As the Majorana CP-phases ($ \rho, \sigma $)  take fixed values  $0^\circ  $ or $ 90^\circ $ at the energy scale $ \Lambda_{\mu \tau} $, depending on their initial choices at $ \Lambda_{\mu \tau} $ one would expect their distinct behavior at low energies due to quantum corrections.
Therefore, we also perform correlation study between the Majorana CP-phases ($ \rho, \sigma $) at low energies.
Besides this, we systematically study the impact of RGE-induced symmetry breaking on  $ 0\nu\beta \beta $ decay  for the different boundary values of $ \rho, \sigma $ in substantial detail. We show our result in a two dimensional conventional Vissani graph in the ($ m_{\rm light}, |\langle m\rangle_{ee} |$) plane \cite{Vissani:1999tu}. We also extend our analysis to the remaining  effective Majorana neutrino masses which may appear in different lepton-number violating (LNV) processes.

We outline this work as follows. In next Sec.~(\ref{sec:SponBrk}), we give a description of the $\mu-\tau$ reflection symmetry and describe its breaking  considering one-loop renormalization-group equations. In Sec.~(\ref{sec:Numer}), we present a detailed set-up of our numerical procedure.
Section~(\ref{sec:Result}) is devoted to our results,  where in Sec.~(\ref{sec:CorrSt}), we perform a correlation study considering the different  parameters which are fixed by the symmetry at $ \Lambda_{\mu \tau} $. Later, in  Sec.~(\ref{sec:NoNuBetaDec}), our results corresponding to $ 0\nu\beta \beta $ decay have been addressed. We discuss various LNV processes other than $ |\langle m\rangle_{ee} | $ in sub-Sec.~(\ref{sec:LNVMasses}).
Finally, we summarize our conclusion in Sec.~(\ref{sec:Conc}). 

\section{Spontaneous Breaking of $ \mu-\tau  $ Reflection Symmetry}\label{sec:SponBrk}

Let us consider the following transformations of the neutrino fields:
\begin{equation}
\nu_{L, e}^{} \leftrightarrow \nu^{c}_{R, e}, ~~~ \nu_{L, \mu}^{} \leftrightarrow \nu^{c}_{R, \tau}, ~~~\nu_{L, \tau}^{} \leftrightarrow \nu^{c}_{R, \mu} \;,
\end{equation}
where $\nu_{L,\alpha}^{}$'s (for $\alpha = e, \mu, \tau$) are the left-handed neutrino fields in the flavor basis, and $\nu_{R,\alpha}^c$'s are the right-handed neutrino charge-conjugated fields of $\nu_{R,\alpha}$'s. These transformations lead to the effective Majorana neutrino mass matrix of the form,
\begin{eqnarray}\label{eq:EffMajMass}
M_{\nu} =   \left( \begin{matrix}\langle m\rangle_{ee} & \langle m\rangle_{e\mu} & \langle m\rangle^{\ast}_{e \mu} \cr
\ast & \langle m\rangle_{\mu\mu} & \langle m\rangle_{\mu \tau} \cr
\ast & \ast & \langle m\rangle^{\ast}_{\mu\mu} \cr
\end{matrix} \right) \;.
\end{eqnarray}
Note that the different entries of the most general Majorana neutrino mass matrix obey the following equalities:
\begin{eqnarray} \label{eq:Mnu_pred}
\langle m\rangle_{ee} = \langle m\rangle_{ee}^* \; , \quad \langle m\rangle_{e\mu}  = \langle m\rangle_{e\tau}^* \; ,\quad \langle m\rangle_{\mu\tau}  = \langle m\rangle_{\mu \tau}^* \; , \quad \langle m\rangle_{\mu\mu}  = \langle m\rangle_{\tau\tau}^* \;.
\end{eqnarray}
The Majorana neutrino mass matrix $M_\nu^{}$ can be diagonalized as $U^{\dagger} M_{\nu}  U^* = m_\nu^d = \mathrm{diag}\{m_1^{}, m_2^{}, m_3^{} \}$. In the standard PDG \cite{Patrignani:2016xqp} parametrization, the unitary mixing matrix $U  =  P_l V P_{\nu}$ can be decomposed as 
\begin{align}\label{eq:pmns}
U =  P_l \left(
\begin{matrix}
c^{}_{12} c^{}_{13} & s^{}_{12} c^{}_{13} & s^{}_{13} e^{-{ i} \delta} \cr 
 -s^{}_{12} c^{}_{23} - c^{}_{12} s^{}_{13} s^{}_{23} e^{{ i} \delta} & c^{}_{12} c^{}_{23} -
s^{}_{12} s^{}_{13} s^{}_{23} e^{{ i} \delta} & c^{}_{13}
s^{}_{23} \cr 
 s^{}_{12} s^{}_{23} - c^{}_{12} s^{}_{13} c^{}_{23}
e^{{ i} \delta} & - c^{}_{12} s^{}_{23} - s^{}_{12} s^{}_{13}
c^{}_{23} e^{{ i} \delta} &   c^{}_{13} c^{}_{23} \cr
\end{matrix} \right) P_{\nu}, \;
\end{align}
where $c^{}_{ij} (s^{}_{ij})$ (for $i < j=1, 2, 3$) stands for $\cos\theta^{}_{ij} (\sin\theta^{}_{ij})$, $ P_l^{} = \mathrm{diag}\{e^{i \phi_{e}},e^{i \phi_{\mu}},e^{i \phi_{\tau}} \}$ contains three unphysical phases which can be absorbed by the rephasing of charged lepton fields, and $ P_{\nu}^{} = \mathrm{diag}\{e^{i \rho},e^{i \sigma},1\}$ is the diagonal Majorana phase matrix. 
Given the form of $M_\nu^{}$ in eq.~(\ref{eq:EffMajMass}), we find  six predictions for the elements of the unitary matrix $U$, namely, 
\begin{equation}\label{eq:prediction}
\theta_{23} = 45^\circ,~~~ \delta=\pm 90^\circ,~~~ \rho,~\sigma = 0 ~~{\rm or}~~ 90^\circ, \phi_{e} = 90^\circ,~~~ \phi_{\mu} = - \phi_{\tau}.
\end{equation}
A detailed description of $ \mu-\tau $ reflection symmetry with the proper phase convention has been discussed in Ref.~\cite{Nath:2018hjx}.

Having discussed the framework of $\mu-\tau$ reflection symmetry, we now proceed to describe the 
breaking of $\mu$-$\tau$ reflection symmetry due to RGE-running in the context of the minimal supersymmetric standard model (MSSM) \footnote{Note that we consider MSSM as our theoretical framework at high energies which can serve as a possible ultraviolet extension of the Standard Model.}.
In this study, we introduce the concerned flavor symmetry at the superhigh energy scale $ \Lambda_{\mu \tau} $ ($ \equiv 10^{14} $ GeV) which is much higher compared to the EW scale  $ \Lambda_{\rm EW} $ ($ \sim 10^{2} $ GeV). Therefore, one must consider  the effect of RGE-running while 
addressing the neutrino oscillation phenomenology at low energies. It is relevant to scrutinize such an effect because, during the RGE-running process, the Yukawa coupling corresponding to $ \mu $ and $ \tau $ may gain a significant difference and later this can severely impact the breaking of $\mu-\tau$ reflection symmetry.
 
The energy dependence of neutrino mass matrix $ M_{\nu} $ is expressed by its RGE-running equation which at the one-loop level
can be written as \cite{Chankowski:1993tx,Antusch:2003kp,Antusch:2005gp,Mei:2005qp}
\begin{eqnarray}\label{eq:RunnEq}
16 \pi^2 \frac{{\rm d}M^{}_{\nu}}{{\rm d}t} = C \left(Y^{\dagger}_l Y^{}_l
\right)^{T} M^{}_\nu + C M^{}_\nu \left(Y^{\dagger}_l Y^{}_l \right) + \alpha M^{}_{\nu} \;.
\end{eqnarray}
Note here that $t$ stands for $\ln (\mu/\mu^{}_0)$ while $\mu$ signifies the renormalization scale,
whereas $C$ and $\alpha$ in the MSSM can be read as
\begin{eqnarray}
&& C = 1 \;,  \alpha \simeq -\frac{6}{5}g^2_1-6 g^2_2+6 y^2_t \;.
\end{eqnarray}
In the diagonal basis of the charged lepton Yukawa coupling matrix, we have $Y^{}_l =
 \mathrm{diag}\{y^{}_e, y^{}_\mu, y^{}_\tau \} $. In the limit $y^{}_{e} \ll y^{}_\mu \ll y^{}_\tau$, one can rule out the contributions of $y^{}_{e}$ and $y^{}_\mu$ compared to $ y^{}_\tau $.
The evolution of $ M^{}_\nu $ due to  the one-loop RGE-running from the high energy scale $\Lambda^{}_{\mu \tau}$ down to $\Lambda^{}_{\rm EW}$ can be expressed as \cite{Ellis:1999my,Chankowski:1999xc,Fritzsch:1999ee}
\begin{eqnarray}\label{eq:RelMuTuEW}
M^{}_\nu(\Lambda^{}_{\rm EW})= I^{}_{\alpha} I^{\dagger}_\tau
M^{}_{\nu}(\Lambda^{}_{\rm \mu \tau}) I^{*}_{\tau} \;,
\end{eqnarray}
where one defines $I^{}_\tau \simeq \mathrm{diag}\{1, 1, 1-\Delta^{}_{\tau}\}$ along with 
\begin{eqnarray}
I^{}_{\alpha} = {\rm exp}\left( \frac{1}{16\pi^2} \int^{\ln \Lambda^{}_{\rm
EW}}_{\ln \Lambda^{}_{\mu \tau}} \alpha ~ {\rm d}t \right) , \hspace{1cm}
\Delta^{}_{\tau} = \frac{C}{16\pi^2} \int^{\ln \Lambda^{}_{\mu \tau}}_{\ln \Lambda^{}_{\rm
EW}}y^2_{\tau}~{\rm d}t \;.
\label{2.3.4}
\end{eqnarray}
Here, $y^{2}_\tau = (1+ \tan^2{\beta}) m^2_\tau/v^2$ in the MSSM, with $ v \approx 174 $ GeV being   the Higgs vacuum expectation value. We notice that $ \Delta^{}_{\tau} $ depends on $ \tan \beta $ which eventually determines the strength of the symmetry breaking pattern. To observe a reasonable amount of deviation from the symmetry, we fix   $ \tan \beta  = 30$ throughout this work.
Further, diagonalization of $ M^{}_\nu $ leads to the approximate expressions of light neutrino masses at low energies as \cite{Huan:2018lzd}
\begin{align}\label{eq:RGEMass}
m_1(\Lambda_{\rm EW}) & \simeq m_1(\Lambda_{\mu \tau})[1 - \Delta_{\tau}(1-c^{2}_{12}c^{2}_{13})]I^{2}_\alpha \;, \nonumber \\ 
m_2(\Lambda_{\rm EW}) & \simeq m_2(\Lambda_{\mu \tau})[1 - \Delta_{\tau}(1-s^{2}_{12}c^{2}_{13})]I^{2}_\alpha \;, \nonumber \\
m_3(\Lambda_{\rm EW}) & \simeq m_3(\Lambda_{\mu \tau})[1 - \Delta_{\tau}c^{2}_{13}]I^{2}_\alpha \;.
\end{align}
Moreover, the different flavor mixing angles at low energies can be given by
\begin{align}\label{eq:RGEAngle}
\theta_{12}(\Lambda_{\rm EW}) & \simeq \theta_{12}(\Lambda_{\mu \tau})  + \dfrac{\Delta_\tau}{2} [s^{2}_{13} (\zeta^{\eta_{\rho}}_{31} -\zeta^{\eta_{\sigma}}_{32} ) + c^{2}_{13} \zeta^{-\eta_{\rho}\eta_{\sigma} }_{21} ]c_{12}s_{12} \;,\nonumber \\ 
\theta_{13}(\Lambda_{\rm EW}) & \simeq \theta_{13}(\Lambda_{\mu \tau})  + \dfrac{\Delta_\tau}{2} [c^{2}_{12} \zeta^{\eta_{\rho}}_{31} + s^{2}_{12} \zeta^{\eta_{\sigma} }_{32} ]c_{13}s_{13} \;,\nonumber \\ 
\theta_{23}(\Lambda_{\rm EW}) & \simeq \theta_{23}(\Lambda_{\mu \tau})  + \dfrac{\Delta_\tau}{2} [s^{2}_{12} \zeta^{-\eta_{\rho}}_{31} + c^{2}_{12} \zeta^{-\eta_{\sigma} }_{32} ] \;,
\end{align}
whereas one can find three CP-phases   at low energies as 
\begin{align}\label{eq:RGEPhase}
\delta(\Lambda_{\rm EW}) & \simeq \delta(\Lambda_{\mu \tau})  + \dfrac{\Delta_\tau}{2} [\dfrac{c_{12}s_{12} }{s_{13}} (\zeta^{-\eta_{\sigma}}_{32} -\zeta^{-\eta_{\rho}}_{31} ) - \dfrac{s_{13}}{c_{12}s_{12} }(c^4_{12}\zeta^{-\eta_{\sigma} }_{32} -s^4_{12}\zeta^{-\eta_{\sigma} }_{31} +  \zeta^{\eta_{\rho}\eta_{\sigma} }_{21} )  ] \;,\nonumber \\ 
\rho(\Lambda_{\rm EW}) & \simeq \rho(\Lambda_{\mu \tau})  + \Delta_{\tau} \dfrac{c_{12}s_{13} }{s_{12}} [s^{2}_{12}  (\zeta^{-\eta_{\sigma}}_{31} -\zeta^{-\eta_{\rho}}_{32} ) + \dfrac{1}{2} (\zeta^{-\eta_{\sigma}}_{32} + \zeta^{\eta_{\rho} \eta_{\sigma}}_{21} ) ] \;,\nonumber \\ 
\sigma(\Lambda_{\rm EW}) & \simeq \sigma(\Lambda_{\mu \tau})  + \Delta_{\tau} \dfrac{s_{12}s_{13} }{2c_{12}} [s^{2}_{12}  (\zeta^{\eta_{\rho} \eta_{\sigma}}_{21} - \zeta^{-\eta_{\sigma}}_{31} ) - c^2_{12} (2 \zeta^{-\eta_{\sigma}}_{32} - \zeta^{-\eta_{\rho}}_{31} - \zeta^{\eta_{\rho} \eta_{\sigma}}_{21} ) ]\;.
\end{align}
Here, $ \eta_{\rho} = \cos2\rho = \pm 1 $ and $ \eta_{\sigma} = \cos2\sigma = \pm 1 $ represents the 
choices of $ \rho $, $ \sigma $ at high energies whereas $ \zeta_{ij} $ with $ i,j = 1,2,3 $ are defined at low energies as  $ \zeta_{ij} = (m_i - m_j)/(m_i + m_j) $.
\section{Numerical Set-up}\label{sec:Numer}
We illustrate our numerical procedure that has been carried out throughout this work in this section. 
By adopting the framework of MSSM, we study the RGE-running effect from $ \Lambda_{\mu \tau} $ down to $ \Lambda_{\rm EW} $. In our numerical analysis, we set $ \tan \beta = 30 $, and the high and low energy boundary scales are fixed to be $ \Lambda_{\mu \tau} = 10^{14}$ GeV and $  \Lambda_{\rm EW} = 91$ GeV, respectively. As the $\mu$-$\tau$ reflection symmetry predicts  maximal value of the atmospheric mixing angle, $ \theta_{23} $, and the Dirac CP-phase, $ \delta $, along with the trivial values of Majorana CP-phases, $ \rho,\sigma $ ($ 0^\circ, 90^\circ $), we consider here four different scenarios, namely, (i) {\bf S1}, $ \rho = 0^\circ,  \sigma = 0^\circ$; (ii) {\bf S2}, $ \rho = 90^\circ,  \sigma = 90^\circ$; (iii) {\bf S3}, $ \rho = 0^\circ,  \sigma = 90^\circ$; and (iv) {\bf S4}, $ \rho = 90^\circ,  \sigma = 0^\circ$. Note that for all  four scenarios, we fix $ \theta_{23} = 45^\circ$ and $ \delta = 270^\circ $, whereas the remaining oscillation parameters (namely, $ \sin^2 \theta_{12}, \sin^2 \theta_{13}, \Delta m^2_{21}, \Delta m^2_{31} $) are scanned over wide ranges with the help of the nested sampling package \texttt{MultiNest} program \cite{Feroz:2007kg,Feroz:2008xx,Feroz:2013hea} at $ \Lambda_{\mu \tau} $ \footnote{The $ \mu - \tau $ reflection symmetry also allows $ \delta = 90^\circ $, however current global-fit seems to favor $ \delta = 270^\circ $ \cite{Capozzi:2016rtj,Esteban:2016qun,deSalas:2017kay}, hence we perform this study for the latter scenario.}. We define the Gaussian-$\chi^2$ function that has been considered in  thenumerical scan as
\begin{equation}
\chi^{2} = \sum_i \dfrac{\left[  \xi_i -  \overline{\xi}_i \right] ^{2}  }{\sigma^2_i} \;,
\end{equation}
where $\xi_i$ represents the neutrino oscillation parameters, i.e., $ \xi_i = \lbrace \theta_{12}, \theta_{13}, \theta_{23}, \delta, \Delta m^2_{21}, \Delta m^2_{31}  \rbrace $ at $ \Lambda_{\rm EW} $. Also, $\overline{\xi}_i$ stands for the best-fit values from the recent global-fit results~\cite{Capozzi:2018ubv}, while $\sigma_i$ signifies the   symmetrized 1$\sigma$ errors. We also define 
the pull for each observable $ \xi_i $ as
\begin{equation}\label{eq:pull}
{\rm pull} (\xi_i) = \dfrac{\left[  \xi_i -  \overline{\xi}_i \right]  }{\sigma_i} \;.
\end{equation}

 The best-fit values and the corresponding 1$\sigma$ errors that we have considered in our numerical simulations~\cite{Capozzi:2018ubv} are $ \sin^2 \theta_{12} = 0.304^{+0.014}_{-0.013},\sin^2 \theta_{13} = 0.0214^{+0.0009}_{-0.0007}, \sin^2 \theta_{23} = 0.551^{+0.019}_{-0.070}, \delta = 1.32^{+0.23}_{-0.18} \pi, \Delta m^2_{21} = 7.34^{+0.17}_{-0.14} \times 10^{-5} {\rm eV}^2, \Delta m^2_{31} = 2.455^{+0.035}_{-0.032} \times 10^{-3} {\rm eV}^2 $. As the current neutrino oscillation data favor normal neutrino mass ordering (i.e., $m_1 < m_2 < m_3$) with more than 3$ \sigma $ C.L. \cite{Capozzi:2016rtj,Esteban:2016qun,deSalas:2017kay} over 
inverted neutrino mass ordering (i.e., $ m_3 < m_1 \sim m_2 $), we concentrate this study considering the former mass ordering. Also, in this study the smallest neutrino mass $ m_1 $ is allowed to vary in the range [0, 0.2] eV.   Based on our numerical analysis, in the next section we describe our results considering the correlation between   $ \theta_{23}, \delta$ as well as $ \rho, \sigma $ at low energies, which arises due to RGE-triggered $ \mu - \tau $ symmetry breaking. Further, we proceed to discuss their impacts on the different Majorana neutrino masses in the $ 0\nu\beta\beta $ decay and LNV processes.

\section{Results}\label{sec:Result}
\subsection{Correlation Study}\label{sec:CorrSt}
By investigating the global-fit of neutrino oscillation data as well as the predictions of the $ \mu - \tau $ reflection symmetry, we choose the initial values of $ \theta_{23} = 45^\circ, \delta = 270^\circ$ at the superhigh energy scale. On the other hand, there is no such preference for the  Majorana phases as there are no experimental results. However, concerned symmetry predicts that both  the  Majorana phases can either be $ 0^\circ $ or $ 90^\circ $. Thus, we perform this analysis for both  choices. The effective Majorana neutrino mass $ |\langle m \rangle_{ee}| $ which can be tested at $ 0\nu\beta\beta $ experiments can behave very differently depending on the values of different Majorana CP-phases. In this context, $ \theta_{13} \rightarrow 0 $ limit is a good approximation to have analytical understanding of $ |\langle m \rangle_{ee}| $ (which we later mention in eq.(\ref{eq:EffMee})) \footnote{For the numerical analysis throughout this work, we consider non-zero value of  $ \theta_{13} $ as given by latest global-fit~\cite{Capozzi:2018ubv}.}.
 We notice that in this limit $ |\langle m \rangle_{ee}| $ depends on CP-phases $ \rho$ and $\sigma $, whereas dependency on $\delta $ can be neglected \footnote{Note that throughout this work we have numerically analyzed RGE-running behavior of all the oscillation parameters, however cancellation among different components of the effective Majorana masses only arise from relative phase factors. Hence, we mainly concentrate on the behavior of different CP-phases.}.
  However, the RGE-running effect of $\delta $ also plays some significant role for the remaining effective Majorana masses. Therefore, we study its behavior considering two less known parameters $ \theta_{23}, \delta $ in the ($ \theta_{23}, \delta $) plane.


\begin{figure}[h!]
\centering
\includegraphics[height=14cm,width=15cm]{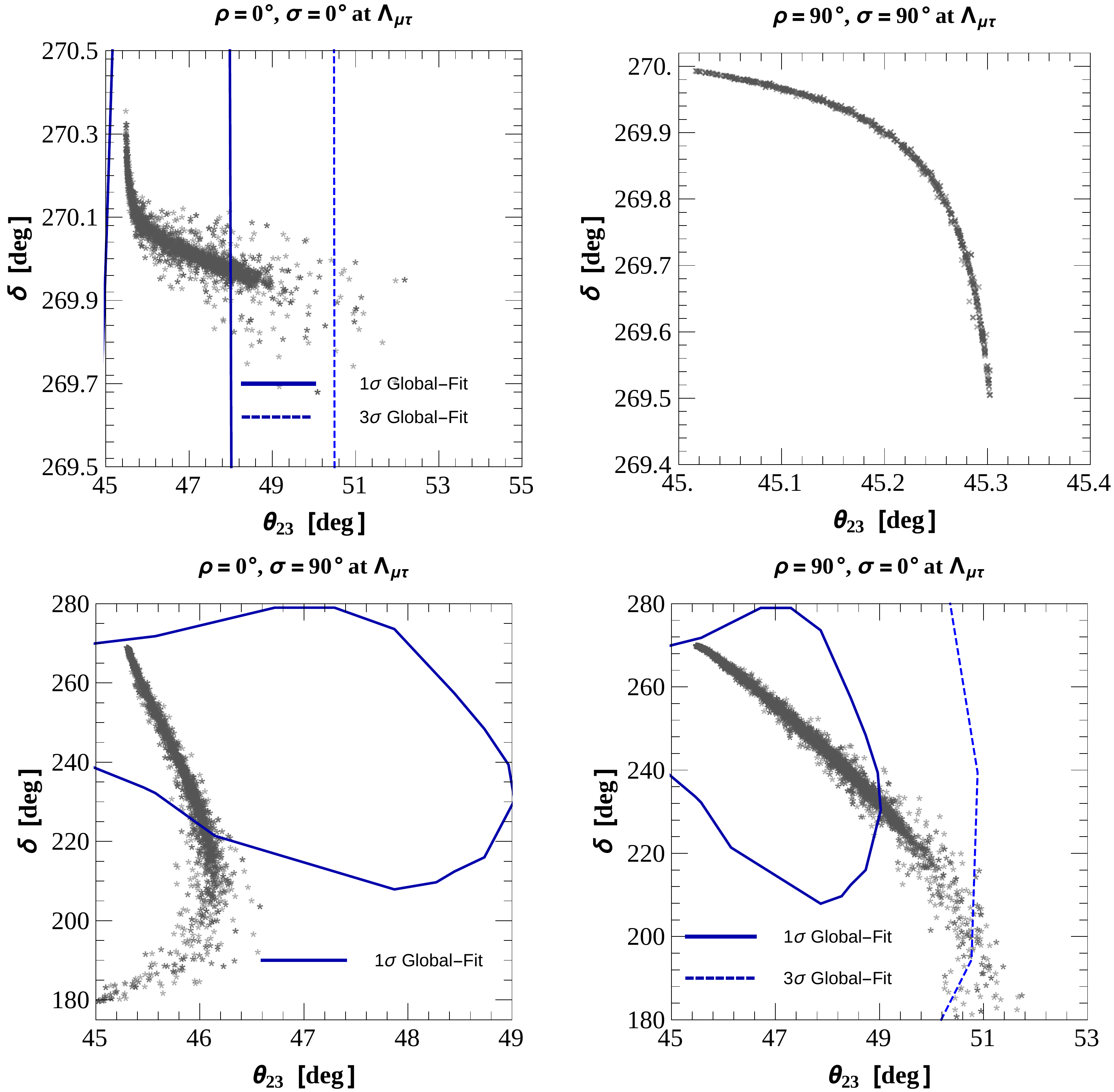}
\caption{\footnotesize Correlation plots between $ \theta_{23}, \delta$ at $ \Lambda_{\rm EW}$. Here, we show four different cases with possible initial values of ($ \rho, \sigma $), as shown by the  plots labels at $ \Lambda_{\mu \tau} $, whereas   $ \theta_{23} = 45^\circ, \delta = 270^\circ$ are adopted at $ \Lambda_{\mu \tau}$ for all cases. Also, the blue contours
in each panel represent the allowed 1$ \sigma $, 3$ \sigma $ area from the global analysis of neutrino oscillation data~\cite{Capozzi:2018ubv}.}
\label{fig:Th23CPCorr}
\end{figure}

In fig.(\ref{fig:Th23CPCorr}), we show the RGE-induced correlation between $ \theta_{23} $ and $ \delta $ at $ \Lambda_{\rm EW}$ for four different possible initial values of $ \rho, \sigma $ at $ \Lambda_{\mu \tau} $ (see figure label for details).  Note that all the gray-scatter points of the figure have $ \chi^{2} < 30 $. Further, we also show allowed 1$ \sigma $ (blue-solid), 3$ \sigma $ (blue-dashed) contours
from the global analysis of neutrino oscillation data~\cite{Capozzi:2018ubv}.
 The results described in fig.(\ref{fig:Th23CPCorr}) are in well agreement with the analytical expressions as mentioned in the third line of eq.(\ref{eq:RGEAngle}) and the first line of eq.(\ref{eq:RGEPhase}) for $  \theta_{23}  $ and $ \delta $, respectively. One can notice from these equations that because of the factor $ \Delta_{\tau} $ which eventually depends on $ \tan\beta $, the difference  ( i.e., $ x(\Lambda_{\rm EW}) - x(\Lambda_{\mu \tau}) \neq 0 $ for $x = \theta_{23}, \delta $) can never be zero. Thus, by inspecting all the panels of   fig.(\ref{fig:Th23CPCorr}), we observe  that in none of the cases scatter curve touches $(  \theta_{23}, \delta) = (45^\circ, 270^\circ)$ which are fixed by the symmetry at high energies.
Further, by inspecting all scenarios, we notice that after the breaking of the symmetry, $ \theta_{23} $   tends to lie always in the higher octant, which is in good agreement with the latest global-fit data. However, the amount of deviation from its initial value is different for different initial boundary values of $ \rho, \sigma $ at $ \Lambda_{\mu \tau} $. 
This can also be understood from the approximate formula of  $  \theta_{23}  $ (see the third line of eq.(\ref{eq:RGEAngle})). As one can notice, the $ \mathcal{O}(\Delta_{\tau}) $ term depends on different CP-phase factors (i.e., $ \zeta^{-\eta_{\rho}}_{31}, \zeta^{-\eta_{\sigma}}_{32} $), which adds a very distinct contribution to $  \theta_{23}  $ depending on the initial options of $ \rho, \sigma $. 
We notice that the current best-fit value of  $ \theta_{23}  \simeq 48^\circ$ can  be reached by the  scenarios {\bf S1} and {\bf S4} as shown by the first and the last panels, respectively. Also in both this scenarios $ \theta_{23} $ can reach values as large as $ 52^\circ $.
%

Considering the RGE-running effect on $ \delta $, we also observe the distinct running behavior of $ \delta $ for four initial options of $ \rho, \sigma $ at high energies. The behavior for distinct deviations remains the same as that explained for $  \theta_{23}  $ and is apparent from the first line of eq.(\ref{eq:RGEPhase}). We find $ \mathcal{O}(1^\circ) $ deviations of $ \delta $ from its maximal value for both scenarios in the top row, whereas deviations of $\mathcal{O}(90^\circ) $ have been noted for both cases in the bottom row. We also notice that deviations of $ \delta $ can reach its current  best-fit value of  $ \delta \simeq 238^\circ$ in both panels of the bottom row.
Furthermore, comparing the top and the bottom row, we observe that the current best-fit value of $  (\theta_{23}, \delta) \simeq (48^\circ, 238^\circ) $ is achievable only for the scenario  {\bf S4}.
By inspecting all the panels of fig.(\ref{fig:Th23CPCorr}), we notice that for the scenarios  {\bf S1} and {\bf S4} there exist some points that lie outside the 3$ \sigma $ contour of the global analysis of neutrino oscillation data as shown by the blue-dashed contour. For {\bf S2}, as the deviation from the exact symmetry due to 
RGE-running is very mild, one can see that the full parameter space lies inside 1$ \sigma $, whereas for {\bf S3} there are some points that fall outside the  1$ \sigma $ contour, but at  3$ \sigma $ all the scatter points lie inside the contour. These also show the consistency of the numerical results compared to the latest global analysis. We emphasize here that as the deviation of $ \delta $ for {\bf S1} and {\bf S2} is very small one cannot see clear contours as obtained from the latest global analysis. However, these contours are visible for the scenarios {\bf S3} and {\bf S4} as one finds large deviations from the symmetry limits.
We also calculate the best-fit points and pull for each observables for all  four scenarios in table (\ref{tab:pull})(see eq.\ref{eq:pull} for the definition of pull).

\begin{table}[t]
\centering
\begin{tabular}{|c | c | c | c | c |}
\hline 
 & {\bf S1} & {\bf S2}   & {\bf S3}  & {\bf S4}\\
Parameter & best-fit \quad pull & best-fit \quad pull & best-fit \quad pull  & best-fit \quad pull  \\
\hline
$\sin^2\theta_{12}$ & 0.3042 \quad 0.015 & 0.3044 \quad 0.029 & 0.3037 \quad -0.022 & 0.3036  \quad -0.029\\
$\sin^2\theta_{13}$ & 0.02138 \quad -0.025 & 0.02142 \quad  0.025& 0.02138 \quad  -0.025 & 0.02139 \quad  -0.012\\
$\sin^2\theta_{23}$ & 0.5520 \quad 0.022& 0.5050 \quad -1.03& 0.5170 \quad  -0.76 & 0.5571 \quad 0.14\\
$\dfrac{\Delta m_{21}^2}{10^{-5}}~\text{eV}^2$ & 7.345 \quad 0.032& 7.343 \quad 0.019& 7.341 \quad 0.006& 7.338 \quad -0.013\\
$\dfrac{\Delta m_{31}^2}{10^{-3}}~\text{eV}^2$ & 2.4560 \quad  0.029& 2.4551 \quad 0.003& 2.4546 \quad -0.029 & 2.4546 \quad -0.029 \\
$\delta /\pi $ & 1.571 \quad 1.224& 1.570 \quad 1.219& 1.28 \quad -0.19& 1.35 \quad  0.15\\
\hline
$\chi^2_{\rm min}$ & 0.77 & 1.80 & 0.63 & 0.03\\
\hline 
\end{tabular}
\caption{The best-fit values and pull  for each observable at low-energies, calculated due to the spontaneous breaking of the symmetry.}
\label{tab:pull}
\end{table}


\begin{figure}[h!]
\centering
\includegraphics[height=14cm,width=15cm]{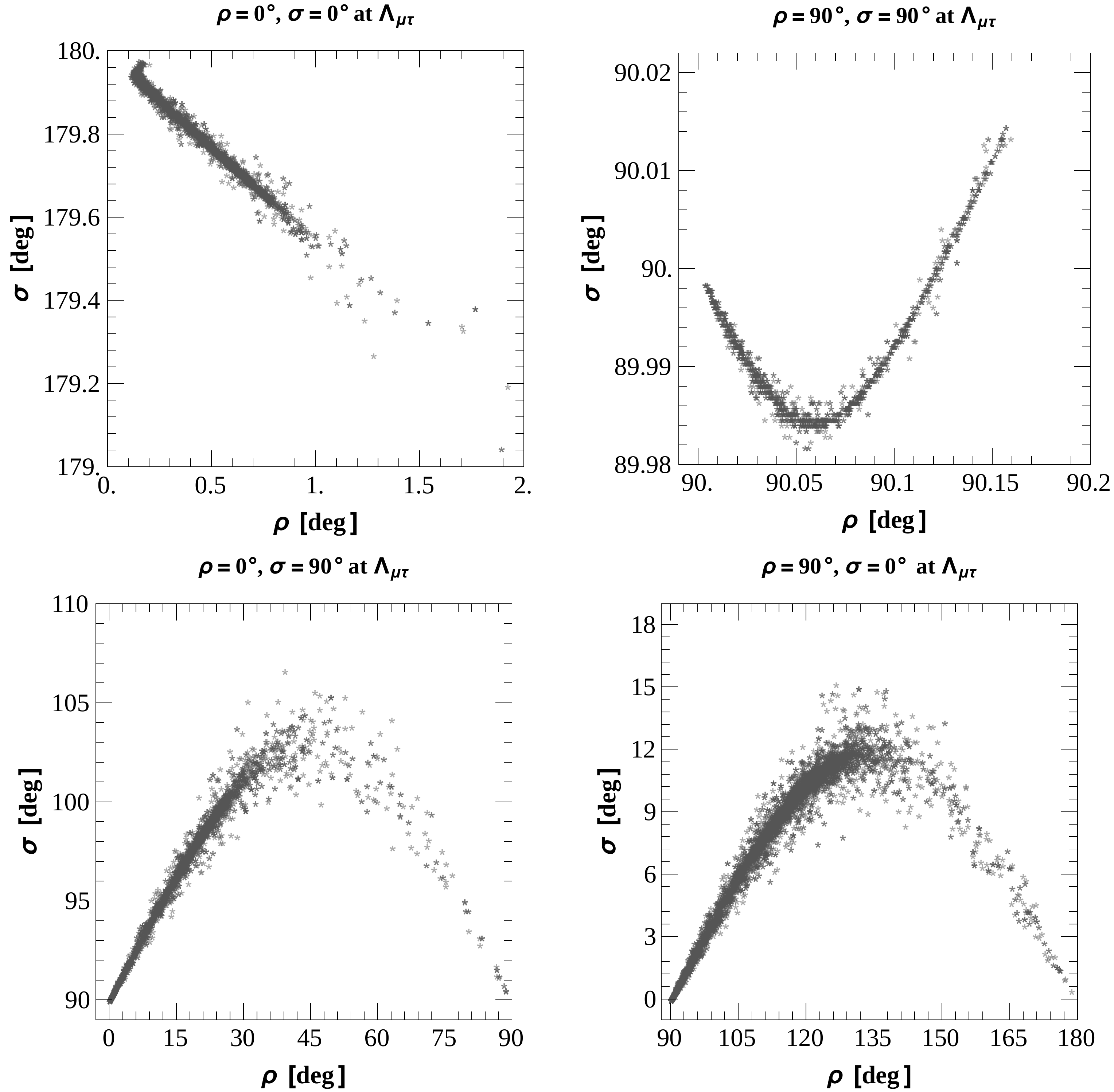}
\caption{\footnotesize Correlation plots between $\rho, \sigma$ at $ \Lambda_{\rm EW}$. We show four different cases with possible initial values of ($ \rho, \sigma $), as shown by the plots labels at $ \Lambda_{\mu \tau} $, whereas   $ \theta_{23} = 45^\circ, \delta = 270^\circ$ are adopted at $ \Lambda_{\mu \tau}$ for all cases.}
\label{fig:RhoSigmaCorr}
\end{figure}

Similarly, in fig.(\ref{fig:RhoSigmaCorr}), we describe the correlation between $\rho, \sigma $ at low energies for four different initial choices of $ \rho, \sigma $ at $ \Lambda_{\mu \tau} $ (see the figure labels for details). By inspecting both of the scenarios in the top row where initial boundary values of $\rho, \sigma $ are chosen to be the same, such as $ 0^\circ $ (left panel) or $ 90^\circ $ (right panel),  we observe a maximum deviation of $ \mathcal{O}(2^\circ) $ for $ \rho $, whereas a deviation of less than $\mathcal{O}(1^\circ) $ has been identified for $ \sigma $.
 However, a noticeable deviation of $ \rho, \sigma $ from their initial values has been observed from the bottom row, where both the Majorana phases take distinct boundary values. We find  deviations of $ \mathcal{O}(90^\circ) $ for $ \rho $, and $ \mathcal{O}(15^\circ) $ for $ \sigma $ from both  scenarios, {\bf S3} and {\bf S4}.
  Similar to the radiative correction of  $ \theta_{23}, \delta$, the numerical pattern of $ \rho, \sigma $ due to RGE-running can also be understood from their analytical results
as given by eq.(\ref{eq:RGEPhase}). We notice very modest correction to $ \rho, \sigma $ from the upper row of fig.(\ref{fig:RhoSigmaCorr}). To explain this, we observe  $ \zeta^{-\eta_{\rho}}_{31}, \zeta^{-\eta_{\sigma}}_{32} $ phase dependency in the  $ \mathcal{O}(\Delta_{\tau}) $ term of eq.(\ref{eq:RGEPhase}) for both $ \rho$ and $\sigma $. One finds a large cancellation between the terms $ \zeta^{-\eta_{\rho}}_{31}, \zeta^{-\eta_{\sigma}}_{32} $  when CP-phases take the same values at high energies. Therefore, we notice 
a very small correction from the upper row, whereas a significant amount of radiative correction
has been observed from the bottom row where both $ \rho$ and $\sigma $ take different initial boundary values, as this leads to very small cancellation among the different phase factors. From this figure, we also notice that due to the correction term $ \mathcal{O}(\Delta_{\tau}) $,  the initial value of  $ \rho, \sigma $ at $ \Lambda_{\mu \tau} $ will never be the same at $\Lambda_{\rm EW}$; in other words, $ x(\Lambda_{\rm EW}) - x(\Lambda_{\mu \tau}) \neq 0 $ for $x = \rho, \sigma$.
 This is apparent from the scenarios {\bf S1} and {\bf S2} as shown in the top row. However, due to the large deviation this behavior is not apparent for  {\bf S3} and {\bf S4} as shown in the bottom row, but our careful zoomed-in analysis shows similar non-zero deviation for both panels of the bottom row.  
From the above discussion, it is now quite apparent that depending on the initial values of the  Majorana phases their behaviors at low energies are very distinct. This may lead to some significant impact 
on the effective Majorana neutrino mass matrix elements.
In the next subsections, we first concentrate our study on $ |\langle m \rangle_{ee}| $ and proceed further to discuss other $ |\langle m \rangle_{\alpha \beta}| $ with  $ \alpha, \beta = e, \mu, \tau $ which may appear in some other lepton-number violating processes.

\subsection{Impact on $ 0\nu\beta\beta $ Decay}\label{sec:NoNuBetaDec}
In this subsection, we scrutinize the impact of the RGE-triggered symmetry breaking effect on neutrinoless double beta ($ 0\nu\beta\beta $) decay  experiments. At the current juncture, $ 0\nu\beta\beta $ decay is the only feasible process which can address the issue of whether the massive neutrinos are the Majorana fermions. This kind of decay process violates lepton number by two-units and the half-life of such a decay process can be written as \cite{Rodejohann:2011mu,Dev:2013vxa}
\begin{equation}
(T^{0\nu}_{1/2})^{-1} = G_{0\nu}|M_{0\nu}(A,Z)|^{2} |\langle m \rangle_{ee}|^{2} \;,
\end{equation}
where $  G_{0\nu}$ stands for the two-body phase-space factor, $ M_{0\nu} $ is the nuclear matrix element (NME), and $|\langle m \rangle_{ee}|$ represents the effective Majorana neutrino mass \footnote{Note that now onwards, we adopt the notation $ |\langle m \rangle_{\alpha \beta}| = |m_{\alpha \beta}| $ for $ \alpha, \beta = e, \mu, \tau $ throughout this work. }. 
The expression of $|\langle m \rangle_{ee}|$ is given by
\begin{equation}
|\langle m \rangle_{ee}| = |m_{ee}| = \left| \sum^3_{i = 1} m_i U^2_{e i} \right| \;,
\end{equation}
where $U$ stands for the PMNS mixing matrix as mentioned in eq.~(\ref{eq:pmns}). In the standard PDG parametrization one can read $ |m_{ee}| $ as
\begin{align}\label{eq:EffMee}
 |m_{ee}| & = | m_1 c^2_{12} c^2_{13} e^{ 2 i \rho} + m_2 s^2_{12} c^2_{13} e^{ 2 i \sigma} + m_3 s^2_{13} e^{- 2 i \delta} | \;, \nonumber \\
  & = | m_1 c^2_{12} c^2_{13}  + m_2 s^2_{12} c^2_{13} e^{ 2 i (\sigma - \rho)} + m_3 s^2_{13} e^{- 2 i (\delta + \rho)} | \;.
\end{align}
Here,  $c^{}_{ij} (s^{}_{ij})$  represents $ \cos \theta_{ij} (\sin\theta_{ij}) $, which are the leptonic mixing angles, while $ \delta $ stands for the Dirac CP-phase  and, $ \rho, \sigma $ signify the Majorana CP-phases. In principle, we can see from the last line of eq.~(\ref{eq:EffMee}) that $  |m_{ee}| $ depends on two effective phases. As the neutrino oscillation data provide information about the mass-squared differences $ \Delta m^2_{21} $, $  \Delta m^2_{31}$ but not about the absolute neutrino masses, we define masses $ m_2 $, $ m_3 $ in terms of the lightest neutrino mass $m_1$ as $ m_2 =\sqrt{m^2_1 + \Delta m^2_{21}}$ and $ m_3 =\sqrt{m^2_1 + 0.5 \Delta m^2_{21} + \Delta m^2_{31}} $ for the normal mass ordering~\footnote{In this study, we define $ \Delta m^2_{21} = m^2_2 - m^2_1 $ and $  \Delta m^2_{31} = m^2_3 - 0.5(m^2_1 + m^2_2)$.}.


\begin{figure}[h!]
\centering
\includegraphics[height=14cm,width=15cm]{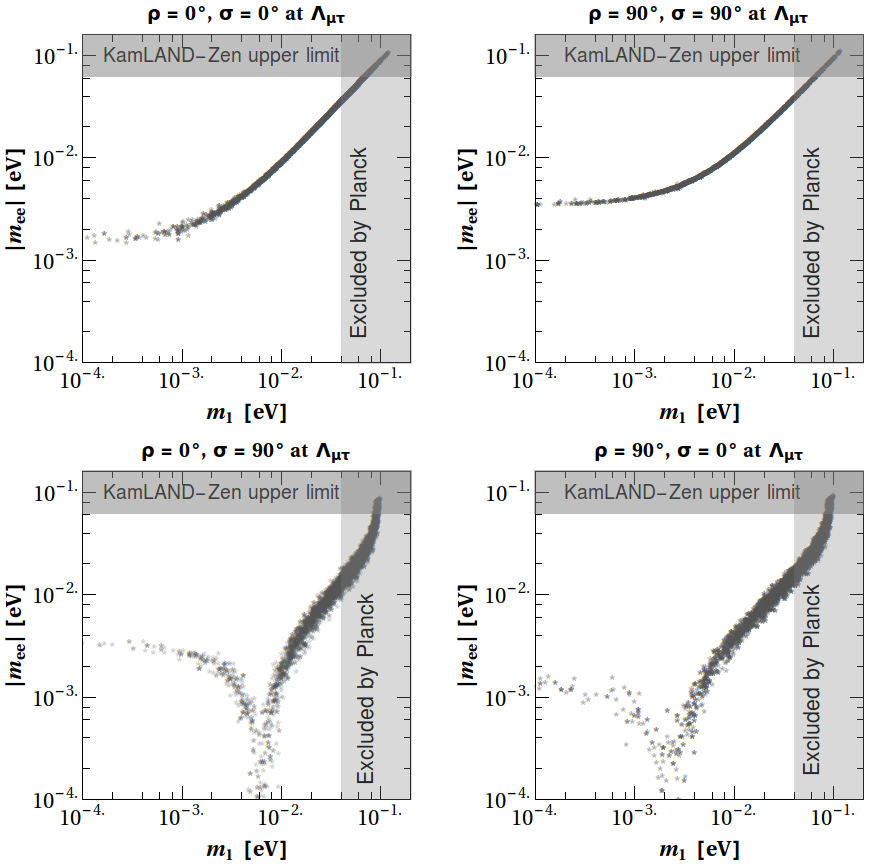}
\caption{\footnotesize Prediction for the effective Majorana neutrino mass $ m_{ee} $ which is involved in $ 0\nu\beta\beta $ decay 
where four panels represent different initial choices of $ \rho, \sigma $. The most stringent upper bound on $ |m_{ee}|  $ from the KamLAND-Zen collaboration is shown by the gray horizontal band. The latest result on lightest neutrino mass is shown by the gray vertical band from the \textit{Planck} Collaboration, which gives $ \sum m_{\nu} < 0.12$ eV at 95\% C.L. }
\label{fig:MeeM1Corr}
\end{figure}

The different experiments which are looking for the signature of neutrinoless double beta ($ 0\nu\beta\beta $) decay  are GERDA Phase II \cite{Agostini:2018tnm}, CUORE \cite{Alduino:2017ehq}, SuperNEMO \cite{Barabash:2011aa}, KamLAND-Zen \cite{KamLAND-Zen:2016pfg} and EXO \cite{Agostini:2017jim}. At present, the most stringent upper bound on the effective Majorana neutrino mass $  |m_{ee}| $ arises from the KamLAND-Zen experiment \cite{KamLAND-Zen:2016pfg}. Their latest results have reported the bound $  |m_{ee}|  < (0.061 - 0.165)$ eV  at 90\% C.L. by taking into account the uncertainty in the estimation of the  nuclear matrix elements.

In fig.(\ref{fig:MeeM1Corr}), we describe our results  of $ 0\nu\beta\beta $ decay for 
 the four boundary values of Majorana phases in the ($m_1,  |m_{ee}| $) plane. These results show the pattern of the effective Majorana neutrino mass $  |m_{ee}| $ due to RGE-triggered breaking of the $ \mu - \tau $ symmetry at low energies.
 The latest results on $ |m_{ee}|  $ from the KamLAND-Zen collaboration is shown by the gray horizontal band in each scenario. On the other hand,  the current results reported by the  \textit{Planck} Collaboration \cite{Aghanim:2018eyx} give $ \sum m_{\nu} < 0.12$ eV (95\%, \textit{Planck} TT, TE, EE + lowE + lensing + BAO). Thus, an upper bound on the lightest neutrino mass has been established as shown by the gray vertical band. Clearly, we notice that depending on the initial values of the Majorana phases one obtains different results of  $ |m_{ee}|  $ at low energies.  

The expression of $ |m_{ee}|  $ at $ \Lambda_{\mu \tau} $ for four different initial values of $ \rho,\sigma $  is given by 
\begin{align}\label{eq:EffMeeAtHE}
 |m_{ee}| (\Lambda_{\mu \tau}) & = | m_1 c^2_{12} c^2_{13}  + m_2 s^2_{12} c^2_{13}  - m_3 s^2_{13}| \;; \quad {\rm for } ~  \rho = 0^\circ,\sigma = 0^\circ \nonumber \;, \\
 & = | m_1 c^2_{12} c^2_{13}  + m_2 s^2_{12} c^2_{13}  + m_3 s^2_{13} | \;; \quad {\rm for } ~  \rho = 90^\circ, \sigma = 90^\circ \nonumber \;, \\
 & = | m_1 c^2_{12} c^2_{13}  - m_2 s^2_{12} c^2_{13}  - m_3 s^2_{13} | \;; \quad {\rm for } ~  \rho = 0^\circ, \sigma = 90^\circ \nonumber \;, \\
 & = | m_1 c^2_{12} c^2_{13}  - m_2 s^2_{12} c^2_{13}  + m_3 s^2_{13} | \;; \quad {\rm for } ~  \rho = 90^\circ, \sigma = 0^\circ\;. 
\end{align}
Here, we kept $ \delta = 270^\circ$ for all cases.

 As the cancellation among the various terms of  $ |m_{ee}|  $ depends on CP-phases, their  behavior at low energies plays a very vital role in understanding the numerical results of fig.(\ref{fig:MeeM1Corr}). In the limit $ \theta_{13} \rightarrow 0 $ which implies $ s_{13} \rightarrow 0, c_{13} \rightarrow 1$, in addition if one finds the deviation of $ \rho, \sigma $ less than $ \mathcal{O}(1^\circ) $ (as shown by the top row of fig.~\ref{fig:RhoSigmaCorr}) then there will not be any significant cancellation among the different components of $ |m_{ee}|  $.  This is apparent from the first two lines of the analytical expressions of eq.(\ref{eq:EffMeeAtHE}). Note that in the numerical analysis, we do not adopt any approximation. From both  panels in the top row of fig.(\ref{fig:MeeM1Corr}), we notice that the minimum of $ |m_{ee}| $  never approaches to zero.
We find that $|m_{ee}|$ can reach around $ \sim 1.5 $ and $ \sim 6 $ meV for $m_1 \rightarrow 0.1 $ meV in the left and right panel, respectively. On the other hand,  the experimentally constrained upper limits of  $|m_{ee}|$  can achieve a value $ \sim 40 $ meV.

We now proceed to elaborate on our results for the second row of fig.(\ref{fig:MeeM1Corr}). Together with the $ \theta_{13} \rightarrow 0 $ limit, if one considers the quasidegenerate scenario (i.e., $ m_1 \approx m_2 \approx m_3 $) then there could be  possible cancellation among the different terms of $ |m_{ee}|  $ as it is apparent from the last two expressions of eq.(\ref{eq:EffMeeAtHE}). This is exactly what we notice from the bottom row where $ |m_{ee}|  \rightarrow 0$ for higher values of $ m_1 $. 
 In other words, one can notice a two-dimensional ``well" in the Vissani graph for the normal neutrino mass ordering in both scenarios {\bf S3} and {\bf S4}. As we notice from the bottom row  of fig.(\ref{fig:RhoSigmaCorr}) where $ \rho, \sigma $ take distinct boundary values, the RGE-running makes a significant contribution to the CP-phases which allows some major cancellation among various components of $ |m_{ee}| $ and leads to $ |m_{ee}| \rightarrow 0$ for some specific range of the lightest neutrino mass. One noteworthy outcome is that,  as the latest neutrino oscillation data 
favor normal neutrino mass ordering and if this turns out to be a true mass spectrum, then this result will play an important role in rulling out or verifying the result of   $ 0\nu\beta\beta $ decay experiments in the standard three-neutrino paradigm. On the other hand, this results will also help us to constrain or determine both the Majorana phases $ \rho, \sigma $.


\subsection{The effective Majorana Neutrino Masses}\label{sec:LNVMasses}

\begin{figure}[h!]
\centering
\includegraphics[height=12cm,width=17cm]{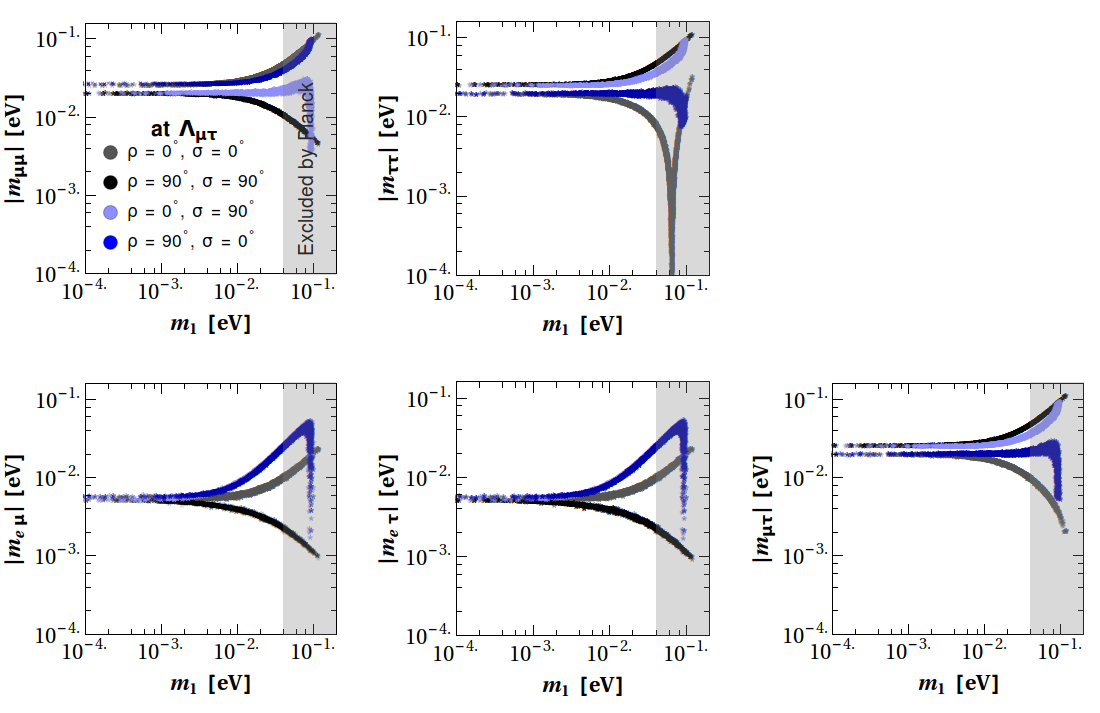}
\caption{\footnotesize Prediction for the effective Majorana masses $ m_{\alpha \beta} $. The gray, black, light-blue, dark-blue patterns show the behavior of $ m_{\alpha \beta} $ for different initial choices of $ \rho, \sigma $ (see figure labels for details). Here, the latest result on lightest neutrino mass is shown by the gray vertical band from \textit{Planck} Collaboration,  which gives $ \sum m_{\nu} < 0.12$ eV at the 95\% C.L.}
\label{fig:MajMassesPlot}
\end{figure}

We extend this study for the remaining  effective Majorana neutrino masses $ |m_{\alpha \beta}| $ for $ \alpha, \beta = e, \mu, \tau $ with $ \alpha = \beta \neq e $. It is important to study $|m_{\alpha \beta}| $ as the number of unknowns involving $|m_{ee}|  $ cannot simply be addressed in $ 0\nu\beta\beta $ decay \footnote{Note that the unknowns in $|m_{ee}|  $ involves the absolute neutrino mass scale, correct mass ordering and the  Majorana CP-phases.}. 
Further, if $ 0\nu\beta\beta $ decay experiments observe null results then one needs to extend their search for other lepton-number violating (LNV) processes to address the Majorana nature of neutrinos. Moreover, the LNV processes involving the effective Majorana neutrino masses can play an important role in the decay rates of $ H^{++} \rightarrow l^{+}_{\alpha}l^{+}_{\beta}  $ and  $ H^{+} \rightarrow l^{+}_{\alpha} \overline{\nu}  $~\cite{Xing:2013woa,Merle:2006du,Rodejohann:2011mu}, in the neutrino-antineutrino oscillation probabilities~\cite{Xing:2013ty,Xing:2013woa}, and in  rare meson decay involving $B$ and $D$ mesons \cite{Atre:2005eb,Rodejohann:2011mu}. 
We list the mathematical form of the effective Majorana neutrino mass matrix elements considering the concerned form of the PMNS mixing matrix,  given by eq.(\ref{eq:pmns}) at high energies as
\begin{eqnarray} \label{eq:LowEnergyElements}
\sqrt{2} |m_{e \mu}| & = & |\overline{m}_1 c_{12}c_{13}(s_{12} - i c_{12} s_{13}) - \overline{m}_2  s_{12}c_{13} (c_{12} + i s_{12}s_{13}) - i m_3 c_{13}s_{13} | \;, \nonumber \\
2 |m_{\mu \mu}| & = & |  \overline{m}_1  (s_{12} - i c_{12}s_{13})^{2}+  \overline{m}_2  (c_{12} + i s_{12}s_{13})^{2} +  m_3 c_{13}^{2} | \;, \nonumber \\
2 |m_{\mu \tau} | & = & | \overline{m}_1  (s_{12}^{2} + c_{12}^{2}s_{13}^{2}) + \overline{m}_2  (c_{12}^{2} +  s_{12}^{2}s_{13}^{2}) -  m_3 c_{13}^{2} | \;, 
\end{eqnarray}
where $ \overline{m}_1 = \pm m_1 $ and $ \overline{m}_2 = \pm m_2 $ with the `$ \pm $' symbol stand for $ \rho, \sigma = 0^\circ $ or $   90^\circ $. Note that the remaining elements of $ M_{\nu} $ satisfy the symmetry equations as mentioned in eq.(\ref{eq:Mnu_pred}).
Fig.(\ref{fig:MajMassesPlot}) illustrates our results, where the impact of spontaneous breaking of the concerned symmetry is introduced on $ |m_{\alpha \beta}| $. We show the  latest \textit{Planck} \cite{Aghanim:2018eyx} bounds on lightest neutrino mass by gray-vertical band at the 95\% C.L.
Note that the current upper bounds on this five  $ |m_{\alpha \beta}| $ can reach multi-MeV to TeV \cite{Atre:2005eb,Rodejohann:2011mu}. One expects the experimental sensitivity of some of the effective Majorana masses would improve to the sub-eV level in the near future.
In fig.(\ref{fig:MajMassesPlot}), the different color patterns like gray, black, light-blue, and dark-blue describe different boundary values of $ \rho, \sigma $, i.e., ($ 0^\circ, 0^\circ $), ($ 90^\circ, 90^\circ $), ($ 0^\circ, 90^\circ $) and ($ 90^\circ, 0^\circ $), at high energies, respectively. 
We notice that these scenarios are indistinguishable for $ m_1 < 0.01 $ eV; in other words, the impact of RGE remains insignificant when $ m_1 $ lies below 0.01 eV.  However, one can distinguish them when $ m_1 $ lies in $ 0.01 \leq  m_1 \leq 0.04 {~\rm eV}$ which also falls in the latest experimentally allowed regime. 
By inspecting all the different cases, we do not notice any significant cancellation among the
 different terms of $   |m_{\alpha \beta}|  $, which may lead to $ |m_{\alpha \beta}|  \rightarrow 0$, except for  $ |m_{\tau \tau }| $   with some specific values of $ \rho, \sigma $ (see the gray curve), but the latest constraint on $ m_1 $ rules out that possibility. However, we find that the effective Majorana masses like  $ | m_{\mu \mu}|, | m_{\mu \tau}|, | m_{\tau \tau}|$  take values as large as $ \sim 80 $ meV for $ m_1 \rightarrow 0 $; on the other hand, channels like $  | m_{e \mu}|, | m_{e \tau}| $ constrain themselves around $ \sim 8 $ meV in the limit $ m_1 \rightarrow 0$. Further, we also notice a very modest impact of RGE-induced symmetry breaking on $  | m_{e \mu}|, | m_{e \tau}| $ as they show almost the same patterns.

\section{Conclusion}\label{sec:Conc}
Numerous neutrino oscillation as well as non-oscillation experiments have opened a new window to probe some intriguing fundamental properties of neutrinos. At the same time, it is also interesting to look for various models which may strengthen our  theoretical understanding. However, in the near future,  more and more data from both oscillation and non-oscillation sectors will help us to verify some definite predictions of different models or rule out some specific models.

In this work, we mainly concentrate on the impact of RGE-induced breaking of the $ \mu-\tau $ reflection symmetry in $ 0\nu\beta\beta $ decay. The experimental result of $ 0\nu\beta\beta $ decay would allow us to establish the Majorana nature of massive neutrinos. Thus, at present, searching for the signal of $ 0\nu\beta\beta $ decay becomes one of the most important task for non-oscillation experiments in neutrino physics. The decay rate of such a process depends on the effective Majorana neutrino mass $  | m_{ee}|$ involving Majorana phases ($ \rho, \sigma $) . Therefore, it is very significant to study $  | m_{ee}|$ which can put bounds on $ \rho, \sigma $. Besides this, the $ \mu-\tau $ reflection symmetry
 predicts $ \theta_{23} = 45^\circ$ and $ \delta = \pm 90 ^\circ$, which are in good agreement with global-fit data along with the trivial values of $ \rho, \sigma $ (i.e., $  0 ^\circ$, $  90 ^\circ$).
However, in order to explain the low-energy neutrino oscillation data, one imposes such symmetry at a superhigh energy scale. In this scenario, it is necessary to consider the RGE-induced 
quantum corrections which naturally breaks the exact symmetry and leads to some interesting aspects of neutrino oscillation data. This also brings some noteworthy correction to the  Majorana phases which has a very significant impact on $  | m_{ee}|$. In this prospect, we make an attempt to explore some salient features of  the effective Majorana neutrino mass $| m_{ee}|$.

 In this paper,  considering the broken scenario, we perform various correlation studies among the parameters (i.e., $ \theta_{23}, \delta, \rho, \sigma$ ) which are fixed by the symmetry at high energies for the normal neutrino mass ordering.
We adopt four different scenarios depending on the initial options of $ \rho, \sigma $ at high energies (i.e.,  (i) {\bf S1}, $ \rho = 0^\circ,  \sigma = 0^\circ$, (ii) {\bf S2}, $ \rho = 90^\circ,  \sigma = 90^\circ$, (iii) {\bf S3}, $ \rho = 0^\circ,  \sigma = 90^\circ$ and (iv) {\bf S4}, $ \rho = 90^\circ,  \sigma = 0^\circ$.). We observe very distinct corrections to both $  \theta_{23}$  and $ \delta $ at low energies depending on the initial values of $ \rho, \sigma $. A deviation of $\sim \mathcal{O}(6^\circ)$ from the maximal value of $  \theta_{23} $  has been noticed for the scenarios {\bf S1} and {\bf S4}, whereas a minute correction of less than $\mathcal{O}(1^\circ)$  has been registered for the scenarios {\bf S2} and {\bf S3}. Similarly, for the Dirac CP-phase $ \delta $, a less than $\mathcal{O}(1^\circ)$  has been observed for the scenarios {\bf S1} and {\bf S2}. However, a large deviation of $\sim \mathcal{O}(90^\circ)$ has been found for the scenarios {\bf S3} and  {\bf S4}. Further, we also notice that only the scenario {\bf S4} is able to reach the latest best-fit values of both $ \theta_{23}, \delta $.
Now, inspecting the correlation between both the Majorana phases ($\rho, \sigma  $), we draw following conclusions. We observe a correction of less than $\mathcal{O}(2^\circ)$ for both $\rho, \sigma  $ when they take same initial values at high energies which are explained by the scenarios {\bf S1} and {\bf S2}. Further, we find a deviation of $\mathcal{O}(90^\circ)$ and $\mathcal{O}(15^\circ)$ for $\rho $ and $ \sigma  $, respectively, from both  scenarios {\bf S3} and {\bf S4}.

Furthermore, by adopting the conventional Vissani graph, we illustrate our result for the effective Majorana neutrino mass in the $ (m_1,| m_{ee}|)$ plane. Depending on the initial choices of 
$\rho, \sigma  $, we observe two distinct behaviors of  $ | m_{ee}| $. Our numerical analysis shows that for the scenarios {\bf S3} and {\bf S4}, the effective Majorana neutrino mass falls into the well where $ | m_{ee}| \rightarrow 0$. This scenario arises because of major cancellation among the different components of $ | m_{ee}| $ and becomes unobservable for $ 0\nu\beta\beta $ decay. However, for the scenarios where the initial options of $\rho, \sigma  $  take the same values, we find no possible cancellation among the terms of $ | m_{ee}| $ as shown by the scenarios {\bf S1} and {\bf S2}. 
We observe  that the magnitude of  $ | m_{ee}|$ varies from 1.5 to 40 meV when the lightest neutrino mass lies in $ 0.1 < m_1 < 0.04~ {\rm eV} $ for the scenario {\bf S1}, whereas $ | m_{ee}|$ takes values from 6 to 40 meV for the same mass range of $m_1$ in the scenario {\bf S2}. Note that we fix the upper limit on the effective Majorana neutrino mass as well as the lightest neutrino mass from the latest experimental bound. This analysis will also help us to constrain or determine both the Majorana phases $ \rho, \sigma $.
Finally, we proceed to discuss the remaining effective  Majorana masses $ | m_{\alpha \beta}|$ with $ \alpha, \beta = e, \mu, \tau $ which may have some interesting consequences for different lepton-number violating channels. In our careful analysis of $ | m_{\alpha \beta}|$ with $ \alpha, \beta \neq e $, we find  no significant cancellation among the different terms of $ | m_{\alpha \beta}|$.
 We observe that the effective Majorana masses like  $ | m_{\mu \mu}|, | m_{\mu \tau}|, | m_{\tau \tau}|$  take values $ \sim 80 $ meV for $ m_1 \rightarrow 0 $, whereas channels $  | m_{e \mu}|, | m_{e \tau}| $ constrain themselves to $ \sim 8 $ meV in the limit $ m_1 \rightarrow 0 $.

\begin{acknowledgements}
Author is  thankful to  Prof.  Zhi-zhong Xing for many useful discussions, insightful comments and careful reading of the manuscript. Author also thanks Mr. Guo-yuan Huang for his help in the numerical analysis and for fruitful discussions. The research work of author was supported in part by the National Natural Science Foundation of China under grant No. 11775231. 
\end{acknowledgements}

\bibliography{mu-tau}
\end{document}